\documentclass[12pt]{article}
\usepackage{authblk}
\usepackage{orcidlink}
\usepackage[square, numbers]{natbib}
\usepackage{hyperref}
\usepackage{url}
\usepackage[margin=1in]{geometry}
\usepackage{multicol}
\usepackage{setspace}
\usepackage{array}
\usepackage{sectsty}
\usepackage{enumitem}

\sectionfont{\fontsize{14}{12}\selectfont}
\subsectionfont{\fontsize{12}{12}\selectfont}

\title{\vspace{-1in}Possibilities for SETI at High Energy\vspace{-0.125in}}
\author[1]{Brian C. Lacki\,\orcidlink{0000-0003-1515-4857}}
\affil[1]{\normalsize Breakthrough Listen, Department of Physics, Denys Wilkinson Building, Keble Road, Oxford OX1 3RH, UK}
\author[2]{Stephen DiKerby\,\orcidlink{0000-0003-2633-2196}\vspace{-0.125in}}
\affil[2]{Department of Physics and Astronomy, Michigan State University, East Lansing, MI 48820, USA}
\date{}

\begin{document}

\maketitle

\vspace{-0.5in}

\section{Why do SETI at high energy?}
\vspace{-0.5em}
Because it is there, of course -- and for all we know, so is evidence that we are not alone. SETI researchers often compare the possible space of observational programs to a haystack, likening technosignatures to the proverbial needle \citep{Wright18}. The high-energy bands are an entire side of the metaphorical barn that has been scarcely explored. 

Astrobiology has been haunted by the question of whether life necessarily resembles Earth life. High-energy astronomy probes some of the most extreme environments out there and, perhaps, some of the most exotic forms of life imaginable. High-energy SETI is about pushing astrobiology to its limits, testing the most fundamental needs of life. 

Postulated technosignatures can require extreme speculation about fanciful technologies that may never be possible, but in the end we have few certain insights into the motivations and forms of extraterrestrial intelligence (ETI). Goals we scarcely comprehend can lead them to places we would otherwise never think to look. There are some positive reasons why technosignatures may lurk in the high end of the spectrum. High-energy radiation emanates from locations with a huge amount of power crammed into a small space, for example. Compact objects like neutron stars could be excellent places for ETIs to operate if they can reach them and brave the hostile environments. 

High-energy SETI has lagged behind the rest of the field even as it has blossomed over the past decade, but the heightened respectability of SETI could spark interest in the coming years. The extant literature has been mainly theoretical, sporadic proposals of possible technosignatures with little observational follow-up, but in many cases, we simply need to take existing data and do the search. Ohio State University is hosting the first workshop on the topic in June 2025.

\section{The panoply of possible high-energy technosignatures}
\vspace{-0.5em}
High-energy emission traces a diverse set of locales and processes. Most relevantly:
\begin{itemize}[topsep=4pt,itemsep=2pt,parsep=1pt]
\item \emph{High-energy radiation comes from nuclear processes}, including reprocessed thermal X-rays, direct gamma-ray line emission from radioisotopes and nuclear de-excitation, and neutrinos.
\item \emph{High-energy radiation comes from relativistic particles}. Diffuse cosmic rays are the most prevalent source, but bulk relativistic flows as found in jets and pulsar wind nebulae also have them. Pions and their resultant electrons, positrons, neutrinos, and gamma rays are ubiquitous in high-energy nucleon interactions. Electrons and positrons create gamma rays too as they interact with radiation, fields, and matter. 
\item \emph{High-energy radiation comes from compact sources} like neutron stars and black holes. The deep gravitational potential of these objects allows for an enormous amount of energy to be extracted from even a small amount of mass.
\item \emph{High-energy radiation comes from transients}, ranging from stellar flares to gamma-ray bursts (GRBs). Relativistic flows are present in the most energetic of these, boosting the energy of photons into the hard X-ray regime and beyond.
\end{itemize}
These principles are guides for the possible types and locations of novel technosignatures.

\subsection{Communication}
\vspace{-0.5em}
The most direct technosignature at high energies is a broadcast with the express goal of communication. X-rays have several possible advantages identified by \citep{Hippke17-XRay}. Their small wavelength minimizes diffraction, potentially allowing tight-beamed transmissions with enormous gains. The high frequencies can sustain higher rates of data transmission, a fundamental consequence of information theory and the expansive phase space afforded to X-rays. The fine time or frequency modulation this allows cannot be probed with current detectors, but we could still detect the underlying burst of emission if it is beamed directly to us. A microsecond flash from a near-earth object or the solar focus of a nearby star would be very anomalous, for example.

Non-directional X-ray communication can be effected by dropping an asteroid onto a neutron star \citep{Corbet97}. When it hits, it releases a burst of energy detectable at interstellar distances. The cosmos also has a number of compact high-energy ``signal lamps''. X-ray binaries (XRBs) are systems with a neutron star or black hole accreting from a donor star, having luminosities of up to $10^5$ suns. Even a kilometer-scale object passing in front of the hotspots of an XRB can easily modulate its luminosity, serving as a technosignature \citep{Corbet97,Imara18}. A subplanetary-scale lens is potentially capable of creating a brief flash visible even in nearby galaxies without any power input of its own \citep{Lacki20}.

Neutrinos are also occasionally suggested as a communication medium \citep{Learned09}. Their weak interaction with matter makes them hard to detect \citep{Hippke18-Messengers}, but also lets them circumvent obstacles. Neutrinos are the only avenue we know for remote transmission into the deep subsurface oceans of icy moons like Europa and Enceladus \citep[cf.][]{Shoji12}, and one of the few in Earth's seas \citep{Huber10}. Even we have experimented with neutrino comm links \citep{Stancil12}, although they remain terribly impractical.

\subsection{Industry}
\vspace{-0.5em}
Industrial activities that could result in high-energy technosignatures include power collection, bulk transport, material synthesis, and scientific experiments. Some possible examples, mostly hypothetical and at the very edge of possibility, are listed in Table~\ref{table:ETIIndustry}.

\begin{table}[h!]
\begin{tabular}[c]{p{5cm}p{9.5cm}c}
\hline
Industrial activity & Possible technosignature & Ref.\\
\hline
Antimatter reactor              & Pionic gamma rays \& neutrinos, positronic gamma rays        & \citep{Harris02}\\
Antimatter rocket               & Annihilation gamma rays, high proper motion                  & \citep{Harris86}\\
Dark matter reactor             & Possible gamma spectral line, neutrinos, other particles     & \citep{Wright14-Search}\\
Nuclear energy usage            & Sub-MeV to MeV antineutrinos, isotopic signatures            & \citep{Hippke18-Messengers}\\
Particle accelerators           & Ultrahigh-energy neutrinos, possible GRB-like transients     & \citep{Lacki15}\\
Primordial black hole reactor & High-energy electromagnetic radiation from evaporation         & \citep{Loeb24}\\
Radioactive ISM tracers         & Gamma-ray lines, nuclides on ocean floor                     & \\
Relativistic shrapnel           & Apparent ultrahigh-energy cosmic rays in atmosphere          & \\
Stellar engineering via neutrino heating & Stray pionic emission                               & \citep{Learned08}\\
Ultrarelativistic craft         & Boosted reflected starlight, interaction gamma-rays          & \citep{GarciaEscartin13}\\
XRB Dyson sphere                & X-ray occultations, too dim in X-rays, infrared excess       & \citep{Imara18}\\
XRB stellar engines             & XRB with unusual proper motion                               & \citep{Vidal24} \\
\hline
\end{tabular}
\caption{Examples of hypothetical high-energy industrial technosignatures.}
\label{table:ETIIndustry}
\end{table}

In order to be detectable at astronomical distances, the industries must be astronomical in scale, perhaps exploiting existing objects like supermassive black holes or XRBs. This is an issue even for demonstrated technosignatures. Our own fission reactors are well known to emit MeV antineutrinos; in fact, they dominate the background at energies below 10 MeV, obscuring diffuse emission from supernovae \citep{Beacom10}. But all of them combined could never even remotely approach the luminosity needed to observe them at interstellar distances. Maybe ETIs resort to nucleosynthesis to fulfill their heavy element needs, but detecting that with neutrinos requires them to burn some solar masses per week \citep[cf.][]{Kato20}. Nuclear explosions are likewise too faint \citep{Hippke18-Messengers}. At present, detecting reactors is realistic only if they are within the Solar System, perhaps one buried in the Moon.

Some conceivable technosignatures have dubious specificity or require implausible circumstances. Gamma-ray spectral lines could trace artificial radioisotope production, just as the Galactic disk is lit up in MeV gamma rays from $^{26}$Al \citep{Diehl06}, but no one has yet suggested why ETIs would produce many planets' worth of these isotopes or visibly expose them to interstellar space. Perhaps they could deliberately release radioisotopes into the interstellar medium (ISM) to trace turbulence and gas motions over long timescales, analogous to our use of dyes in seawater. Other things we can posit will be impractical or impossible. Wormholes and warp bubbles could display high-energy markers \citep[cf.][]{Bambi21}, but creating them is a long shot. Cosmic strings, useful they may be, have never been found. We can speculate about engineering stars arbitrarily, but could anyone actually \emph{do} that? We do not know which remote possibilities will bear out. 

Yet some technologies have practically inevitable signatures if they do exist. X-ray sources enclosed in Dyson spheres are necessarily dimmed and accompanied by waste heat, simply from energy conservation \citep{Dyson60}. Any particle accelerator probing Planck-scale physics is likely to be luminous because of an overwhelming background of mundane particle interactions \citep{Lacki15}. A relativistic artificial grain that survives the perilous journey to Earth \citep{Hoang15}, be it shrapnel or a microscopic probe, triggers pionic showers in the atmosphere with total energy similar to an ultrahigh-energy cosmic ray.
        
\subsection{Habitat}
\vspace{-0.5em}
What are the limits of life, broadly defined? At the very least, complex processes require a thermodynamic gradient to feed them. In his reflections on the future of the cosmos, Dyson suggested that this is the only absolute requirement, and that long after the stars have gone out, life could still thrive in the chilly atmospheres of cooled compact objects \citep{Dyson79}. A contemporary test of this admittedly extreme idea might be found with today's compact objects. The accretion hotspots of XRBs have some of the greatest sustained power densities around in the contemporary universe. If thermodynamics really is the only prerequisite factor for complexity and ETIs can withstand the incredibly hostile environments, they may find the energy graidents in XRBs attractive \citep{Vidal16}. Nuclear-based life on the surfaces of neutron stars is another similarly extravagant possibility. Any intelligence that evolves there is likely trapped by the extreme gravitational potential, and thus could only communicate remotely with high-energy radiation.

Far more prosaically, high-energy radiation can signpost various threats to conventional habitability like stellar flares, supernovae, and GRBs. ETIs may attempt to suppress or deflect them if they can.

\section{Potential avenues forward}
\vspace{-0.5em}
High-energy SETI by and large must be a commensal effort for the foreseeable future. Dedicated programs will only be feasible after much further investigation. At this stage, our efforts will be like those of the early radio and optical SETI pioneers who developed methods and infrastructure that took decades to grow into the robust subfield it is today. An even more basic reason for commensal studies is the difficulty in building optics for some kinds of radiation. Because we cannot make neutrino lenses, every neutrino detector is sensitive to large sky areas, making it a commensal SETI facility.

A high-energy SETI program is a search for anomalies. Machine learning has grappled with the problem of finding general outliers in a dataset. These methods have already been applied to conventional SETI and there is no reason they could not be applied to high-energy datasets. Of course, far more likely than detecting an alien signal is finding a new astrophysical phenomenon, but this in itself would be a valuable find.

Most attempts at communication and some industrial activities would probably look like a brief burst of radiation, or possibly a short dip in something bright. There is already a wealth of data to be mined: from X-ray images and timing data, to GRB triggers, to years of gamma-ray observations, to searches for neutrino bursts. Nonetheless, they may be very rare or faint. Technosignatures can have several distinct features. X-ray lenses with chromatic aberration produce a peculiar spectrophotometric signature when they transit bright sources \citep{Lacki20}. Pionic signatures from antimatter annihilation would have an unusually narrow gamma-ray spectrum, unlike the more common power laws \citep{Harris02}. At the limits of possibility, the bulk relativistic flows of artificial GRB-like events might serve as the engines of particle accelerators, possibly marked with unusually low luminosities or extreme Lorentz factors. Communication attempts may have unusual timing patterns. We should also be vigilant about \emph{where} transients occur \citep{Lacki21}. Bright X-ray bursts coming from nearby stars or Solar System objects are suspect. Sources with high proper motion and acceleration are especially noteworthy.

Technosignatures do not just include individual events or objects, but statistical properties of populations. XRBs are easily detected in nearby galaxies, following known correlations \citep{Lehmer10}. If ETIs favor them and are capable, there may be galaxies where they are all enclosed by Dyson sphere-like structures, or where even stellar evolution has been manipulated to overproduce them. These would stand out, having too many or too few discrete X-ray sources for their mass and star-formation rate, and more generally in total X-ray emission \citep[cf.][]{Lacki24-ETIPops2}.

The case of Dyson spheres reminds us that we should also be alert for objects that are too \emph{quiet} in high-energy radiation. To speculate freely, perhaps ETIs develop something along the lines of ``helio-engineering'' that suppresses stellar flares to maintain the habitability of planets around otherwise active red dwarfs, which might show up as a lack of coronal X-rays.

The Solar System could host high-energy technosignatures, alongside natural phenomena like solar flares, the Jovian radiation torus, and terrestrial gamma flashes. Heliophysics missions monitor the Sun and its surroundings in X-rays and gamma rays \citep{Smith04}. Particle instruments on planetary probes are capable of detecting high-energy transients \citep{Hurley13} and could also be relevant. In the far future, samples from deep below the Moon's surface could act as ``paleodetectors'' for high-energy events \citep{Baum24}. Astrophysics, heliophysics, and planetary sciences may all overlap with astrobiology in high-energy SETI. 

All considered, this branch of SETI has even more uncertainties than conventional SETI and astrobiology, particularly in its current early stage. While it is unlikely that a particular search will turn up anything, high-energy SETI is a subfield with the potential to dawn in the next years.

\subsubsection*{References}
\vspace{-0.5em}
\bibliographystyle{smol}
\bibliography{HighEnergySETI_DARES2025_arXiv}

\end{document}